\documentclass[11pt]{article}

\usepackage{enumerate,natbib} 

\newcommand{\blind}0
\def\spacingset#1{\renewcommand{\baselinestretch}{#1}\small\normalsize}   

\usepackage{color,charter,graphicx,natbib,fancyhdr,bibentry,enumitem,eso-pic,here,setspace,multirow,amsmath,amssymb,amsfonts,subfigure,mathdots, amsmath, sidecap}
\usepackage[svgnames,x11names]{xcolor}
\usepackage[pdftex]{hyperref} 
\hypersetup{
colorlinks,%
linkcolor=MidnightBlue,  
urlcolor=DarkSlateGray,   
citecolor=RoyalBlue2  
}
\sidecaptionvpos{figure}{c}

\def\blu{\color{RoyalBlue4}}

\def\cM{{\cal M}} \def\cD{{\cal D}}
 
\def\btheta{\pmb{\theta}}
\def\bomega{\pmb{\omega}}
\def\bphi{\pmb{\phi}}

\def\bmu{\pmb{\mu}}
\def\bgamma{\pmb{\gamma}}
\def\blambda{\pmb{\lambda}}

\def\bB{\pmb{\Psi}}
\def\bbeta{\pmb{\beta}}

\def\bA{\textbf{A}}
\def\bB{\textbf{B}} 
\def\bC{\textbf{C}}
\def\bF{\textbf{F}}
\def\bR{\textbf{R}}

\def\bG{\textbf{G}} 
\def\bQ{\textbf{Q}}
\def\bS{\textbf{S}}
\def\bW{\textbf{W}}

\def\bb{\textbf{b}}
\def\bff{\textbf{f}}
\def\bh{\textbf{h}}
\def\by{\textbf{y}}
\def\bz{\textbf{z}}
\def\bm{\textbf{m}}
\def\ba{\textbf{a}}

\def\bzero{\textbf{0}}
\def\bone{\textbf{1}}
\def\tr{\textrm{tr}}

\begin{document}

	\if0\blind
	{    
		\title{Bayesian Computation in Dynamic Latent Factor Models}
		\author{Isaac Lavine\thanks{The research reported here was developed while Isaac Lavine was a PhD student in Statistical Science at Duke University, and was partially supported by 84.51$^\circ$, 100 West 5th Street, Cincinnati, OH 45202.}
			\hspace{.2cm}\\
			SIG/Susquehanna International Group\\  401 City Avenue,  Bala Cynwyd, PA 19004 
			\bigskip \\
			Andrew Cron\\
			84.51$^\circ$\\ 100 West 5th Street, Cincinnati, OH 45202
			\bigskip \\ 
			and 
			\bigskip \\ 
			Mike West \\
			Department of Statistical Science, Duke University\\ Durham NC 27708-0251
			}
		\maketitle
	} \fi
	\if1\blind
	{     
		\bigskip
		\bigskip
		\bigskip
		\begin{center}
			{\LARGE\bf Title}
		\end{center}
		\medskip
	} \fi
%
%
	
\bigskip
\begin{abstract}
	Bayesian computation for filtering and forecasting analysis is developed for a broad class of dynamic models.  The ability to scale-up such analyses in non-Gaussian, nonlinear multivariate time series models is advanced through the introduction of a novel copula construction   in sequential filtering of coupled sets of dynamic generalized linear models. The new copula approach is  integrated into recently introduced multiscale models in which univariate time series are coupled via nonlinear forms involving dynamic latent factors representing cross-series relationships.   The resulting methodology offers
dramatic speed-up in online Bayesian computations for sequential filtering and forecasting in this broad, flexible class of multivariate models.  Two examples 
in nonlinear models for very heterogeneous time series of non-negative counts demonstrate 
massive computational efficiencies relative to existing simulation-based methods,  while defining similar filtering and forecasting outcomes. 
\end{abstract}

\noindent%
\vfill

\setcounter{page}0\thispagestyle{empty}

\newpage
\spacingset{1} 

%

\section{Introduction}

Scaling dynamic modeling of multivariate time series is increasingly addressed with Bayes\-ian hierarchical and latent factor state-space approaches. Traditional hierarchical 
methods~\citep[e.g.][]{GamermanMigon1993,Cargnoni1997,FerreiraGamermanMigon1997} couple 
univariate series into dependent multivariate structures via stages of conditional models. Dynamic latent factor models of various 
forms~\citep[e.g.][]{LopesCarvalho07,DelNegro2008,Carvalho11,NakajimaWest2017BJPS,McAlinnEtAl2017,ZhouNakajimaWest2014IJF}
relate sets of univariate time series via underlying processes impacting collectively but in series-specific ways. Such approaches are now very commonly applied, but their Bayesian analyses are computationally demanding and usually involve  simulation-based methods including, in particular, Markov chain Monte Carlo~\citep{Ferreira2006,Carvalho2010, kastner2017,McAlinnEtAl2017, McAlinnWest2017bpsJOE,NakajimaWest2013JFE}.   This reliance on intense simulation methods limits the ability to 
scale in terms of time series dimension, and especially impacts on sequential analysis and forecasting with incoming/streaming data. 
Among major recent innovations to address this,   the decouple/recouple modeling strategy~\citep{West2020Akaike} has opened the path to substantial improvements in computational efficiency of sequential analysis in state-space models. Nevertheless, implementation has, to date,  relied on substantial use of Monte Carlo simulation, inherently limiting scalability. This is a main  challenge to computational statistical science in the important area of dynamic modeling for sequential learning, monitoring and forecasting with increasingly high-dimensional time series.

The current paper addresses these challenges.  The two main, interlinked  advances are as follows. First, we define extensions of the traditional Bayesian analysis in univariate dynamic generalized linear models (DGLMs) to a broad, flexible class of multivariate models. This builds on the multiscale approach for dynamic latent factor modeling to induce and represent multivariate dependencies across series and their potential changes over time.   Second, the computational advances are partly based on novel theoretical developments using copula constructions to represent time-specific relationships across series with univariate marginal forecast distributions constrained by traditional univariate DGLMs.   In these senses, the developments here theoretically extend standard univariate DGLM analysis to flexible classes of multivariate models, enabling computations to dramatically improve upon existing simulation-based methods.
 
Section \ref{section-dglm}  summarizes the context, notation and aspects of traditional analysis of univariate DGLMs, and then discusses multivariate models in which
dynamic latent factor processes define cross-series relationships.    This Section introduces a new extension of the traditional Bayesian analysis of DGLMs to this class of dynamic latent factor structures.  Important sub-classes of these models are conditionally Gaussian models, as well as classes of nonlinear models for discrete time series of non-negative counts.  Problems in scaling time series monitoring and forecasting for count time series are one main motivating context, and the model class discussed here includes  recent models for such problems in both large-scale forecasting~\citep{Aktekin2018,BerryWest2018DCMM, BerryWest2018TSM} and 
dynamic network flow monitoring~\citep{ChenETALdynets2016JASA,ChenBanksWest2018}. Such contexts can involve very high-dimensional time series and the standard use of  simulation analysis  restricts  scalability.   In addition to forward filtering, many applications involve forecasting multiple steps ahead 
at each time point, and successive revision of multi-step and path forecasts as time progresses, which adds to computational demands.

Our main methodological innovation develops a copula method to extend traditional  analysis of univariate DGLMs to the multivariate setting. 
Section \ref{analytic-inference} gives details on copula constructions and their embedding in sequential  analyses of sets of DGLMs that are coupled via dynamic latent factor processes.  This also includes the use of the copula method in multi-step ahead and path forecasting analysis. Sections \ref{section-salesforecasting} and \ref{section-networkflow} demonstrate the methodology and its comparison with the existing simulation-based analysis in applied monitoring and forecasting contexts with count time series.  The examples highlight the generation of equivalent analysis and forecasting performance, but with very major computational efficiencies.  This forcefully demonstrates the main advances in statistical computation;  the  new copula-based modeling methodology defines the potential for orders-of-magnitude advances in the ability to scale-up Bayesian filtering and forecasting in this rich class of multivariate/multiscale dynamic models. Appendices and supplements provide additional technical details and comprehensive code, with examples, for readers interested in  applying or extending the computational methodology.

\section{Dynamic Modeling, Forecasting, and Latent Factors} \label{section-dglm}
\subsection{Univariate DGLMs} \label{section-dglm-algos} 
Structure and notation for the standard and widely-used class of dynamic generalized linear models (DGLMs) follows~\citet[][chapter 14]{WestHarrison1997}. We have a set of $N$ time series $y_{i,t}$ in equally-spaced time $t;$ the series $i=1{:}N$ are modeled conditionally independently given model   parameters.  With $\cD_t$ denoting all information up to time $t$, the conditional distribution of $y_{i,t}$ is of exponential family form
\begin{equation} \label{eqn-exponential-family}
p(y_{i,t}| \mu_{i,t}, \tau_{i,t}, \cD_t) = b(y_{i,t}, \tau_{i,t}) \exp{ \left[ \tau_{i,t}(y_{i,t} \mu_{i,t} - a(\mu_{i,t})) \right]}, \quad i=1{:}N, \ \ t=1,2,\ldots, 
\end{equation} 
with natural parameter $\mu_{i,t}$ and precision parameter $\tau_{i,t}$. The former maps to the linear predictor
$\lambda_{i,t} = g(\mu_{i,t})$ with a specified link function $g(\cdot).$ The state-space model involves dynamic regression and Markov evolutions of   state vectors (series-specific, time-varying regression parameter vectors) $\btheta_{i,t}$ via 
\begin{equation} \label{eqn-state-space}
\lambda_{i,t}  = \bF_{i,t}' \btheta_{i,t}  
\qquad\textrm{where}\qquad 
\btheta_{i,t} = \bG_{i,t} \btheta_{i,t-1} + \bomega_{i,t} \quad\textrm{with}\quad \bomega_{i,t}  \sim \left[\bzero, \bW_{i,t}  \right]
\end{equation}
based on quantities as follows: $\bF_{i,t}$ is a known vector of constants and/or  predictor variables;
$\bG_{i,t}$ is a known state evolution matrix;  $\bomega_{i,t}$ is a stochastic state innovation vector with mean $\bzero$ and variance matrix $\bW_{i,t};$ the innovations  $\bomega_{i,t}$ are independent over time $t$ and across series $i$, and  
conditionally independent of all past states $\btheta_{*,s}$ for $s<t; $ each state evolution variance matrix $\bW_{i,t}$ is defined through discount factors on state vector components~\citep[e.g.][chapter 6]{WestHarrison1997}.

Critical to the interest in scaling to many time series is the central role of the fast, efficient and effective DGLM algorithm that exploits coupled variational Bayes' (VB) and linear Bayes' (LB) steps at each time point~\citep[e.g.][]{migon1985application,West1985a,Triantafyllopoulos2009}. With a main focus on monitoring and forecasting analyses,   the {\em VBLB algorithm} for DGLMs has for many years been a cornerstone in implementations. Of particular relevance here, VBLB features centrally in more recent work on scalable multivariate models for count time series of various kinds~\citep{ChenETALdynets2016JASA,ChenBanksWest2018,BerryWest2018DCMM, BerryWest2018TSM} which exploit Bernoulli, binomial and Poisson DGLMs as well as Gaussian cases. In the latter case, assuming normality of the innovations $\bomega_{i,t}$ leads, of course, to the VBLB algorithm defining the usual Kalman filter-based analysis.  The distributional form for $\bomega_{i,t}$ is rarely of substantive interest and, in linear models, a normal assumption is simply convenient. In  non-normal contexts-- while practical circumstances predicate on innovation distributions that are unimodal and symmetric about zero-- the role of innovations is simply to add random perturbations in the evolution. Partly reflecting this, DGLM analysis is semi-parametric and relies only on first and second order moments. Specific details are central to the extensions to latent factor models of this paper, so are summarized here in the univariate DGLM setting. 

\paragraph{VBLB for DGLMs.}   The univariate analysis of one series $i$ has the following features. 
\begin{itemize} \itemsep-3pt 
\item At time $t-1,$ historical information $\cD_{t-1}$ is summarized by the current posterior mean vector and variance matrix for the state vector  as
 $\btheta_{i,t-1} | \cD_{t-1} \sim [ \bm_{i,t-1},\bC_{i,t-1} ].$   
 \item Evolving to time $t$ this implies prior (to time $t$) moments  $\btheta_{i,t} | \cD_{t-1} \sim [ \ba_{i,t},\bR_{i,t} ]$   with $\ba_{i,t} = \bG_{i,t} \bm_{i,t}$ and 
 $\bR_{i,t} = \bG_{i,t} \bC_{i,t-1} \bG_{i,t}' + \bW_{i,t}.$  This implies prior moments for the linear predictor $\lambda_{i,t}|\cD_{t-1} \sim [f_{i,t},q_{i,t}]$ where $f_{i,t}=\bF_{i,t}'\ba_{i,t}$ and 
 $q_{i,t}=\bF_{i,t}'\bR_{i,t}\bF_{i,t}.$ 
 \item The VB step defines the unique {\em conjugate}   prior $p(\mu_{i,t}|\cD_{t-1})$ consistent with the values $[f_{i,t},q_{i,t}].$ This yields the 
 $1-$step forecast distribution $p(y_{i,t}|\cD_{t-1})$ in analytic form. 
 \item  Observing $y_{i,t}$ updates to the conjugate posterior $p(\mu_{i,t}|\cD_t).$
 This implies posterior moments for the linear predictor $(\lambda_{i,t}|\cD_t) \sim [ g_{i,t},p_{i,t}].$ 
 \item The LB step defines the filtering update for the moments of the state vector via  $\btheta_{i,t} | \cD_t \sim [ \bm_{i,t},\bC_{i,t} ]$ where
 $\bm_{i,t} = \ba_{i,t} + \bA_{i,t}(g_{i,t}-f_{i,t})$ and $\bC_{i,t} = \bR_{i,t} -\bA_{i,t}\bA_{i,t}'(q_{i,t}-p_{i,t})$ with {\em adaptive vector} $\bA_{i,t}=\bR_{i,t}\bF_{i,t}/q_{i,t}.$   
 \item Time index $t$ updates by 1, and the filtering and forecasting cycle repeats. 
 \end{itemize} 
 
 \paragraph{Multi-step Ahead and Path Forecasting.} 
At time $t$ given $\cD_t,$ , consider   forecasting over  times $t+1{:}t+k$ for some integer $k>1$. This is traditionally based on recursive use of the VBLB algorithm into the future, coupled with direct simulation. 
An important distinction is that between {\em marginal} and {\em joint}  forecasting.  With $\btheta_{i,t}|\cD_t \sim [\bm_{i,t},\bC_{i,t}],$ recursing the state evolution equation into the 
future trivially evaluates the mean and variance matrix of each future state vector, and hence the mean and variance of  $\lambda_{i,t+h}|\cD_t$ for each $h=1{:}k.$ Applying the VB step 
defines a conjugate form for $p(\mu_{i,t+h}|\cD_t)$ and hence the {\em marginal} forecast distribution $p(y_{i,t+h}|\cD_t)$ is fully available.  In contrast, {\em joint} forecasting aims to 
evaluate  aspects of $p(y_{i,t+1},\ldots,y_{i,t+k}|\cD_t)$ more comprehensively, exhibiting the dependencies over time as well as marginal, time-specific aspects. {\em Path} forecasting 
based on joint distributions is critical in many areas. In forecasting consumer sales or demand, and in revenue or index forecasting in commerce and economics, interest often lies in implied 
forecasts for aggregates, cumulative totals and/or percentages over a period of time, as well as in the  $y_{i,t}$ themselves. This inevitably uses forward simulation, as follows: 
\begin{enumerate}  \itemsep-3pt 
\item Using VB to define the conjugate-based $1-$step ahead predictive distribution, sample a \lq\lq synthetic outcome'' $y_{i,t+1}^{(s)} \sim p(y_{i,t+1}|\cD_t)$.
\item Conditional on this sampled value, apply LB to update the moments of $\btheta_{i,t+1} | \cD_t, y_{i,t+1}^{(s)}$.
\item Evolve the state vector moments through the evolution equation to time $t+2,$ apply VB and repeat with time index increased by 1, so generating $y_{i,t+2}^{(s)} \sim p(y_{i,t+2}|\cD_t,y_{i,t+1}^{(s)})$.
\item Repeat steps 1 and 2 with time index increased to $t+3,$ and continue to time $t+k$. 
\end{enumerate}  The result is a single Monte Carlo sample  of the  path  $\{ y_{i,t+1}^{(s)},\ldots,y_{i,t+k}^{(s) }\}.$    Independent replicates define a full Monte Carlo sample that can be summarized as desired.  
More details appear in context in the Appendix. See also chapter 14 of ~\cite{WestHarrison1997} and in example in~\cite{BerryWest2018DCMM}.   Clearly, with higher $k$ and when applied collectively across a large number of series $N,$ this becomes increasingly computationally expensive.  One application of our new methodology, below, addresses this.

\subsection{Dynamic Latent Factors and Multiscale Modeling} \label{section-multivariate-model}
 
The methodological and computational developments of this paper relate to multivariate models in which the set of $N$ time series are linked via dynamic latent factor processes.  Model $\cM_i$ for series $i \in 1{:}N$ is a DGLM of eqns.~(\ref{eqn-exponential-family},\ref{eqn-state-space}), now {\em conditional} on the values of a vector latent factor process $\bphi_t$ at all times $t.$   Dependencies across series are induced as a result of the latent factor process.   
In our  general setting we adopt the multiscale modeling perspective in which the  shared information across series through the factor process $\bphi_t$ is inferred from an independent, external model labelled $\cM_0$.    The overall multivariate model is then  
\begin{align} \label{eqn-multivariate-model}
\cM_i: & \quad \textrm{Eqns.}~(\ref{eqn-exponential-family},\ref{eqn-state-space}) \quad\textrm{with}\quad 
	\btheta_{i, t} =	\begin{pmatrix}	\bgamma_{i,t} \\	\bbeta_{i,t}	\end{pmatrix}\quad \textrm{and}\quad
	\bF_{i,t}= \begin{pmatrix}	\bh_{i, t} \\	\bphi_t  \end{pmatrix},  \quad \textrm{for}\ \ i=1{:}N;
\\
\cM_0: &  \quad \bphi_t \sim p(\bphi_t|\cM_0,\cD_{t-1}). 
\end{align}
Here  $\bbeta_{i,t}$ is the series $i$ state subvector corresponding to the latent factor process $\bphi_t$, and the latter arises from the specified external or prior model $\cM_0.$    The rich literature on dynamic latent factor models, mostly in conditionally linear, Gaussian settings, has defined substantial advances in time series forecasting in recent years~\citep[e.g.][]{LopesCarvalho07,DelNegro2008,Carvalho11,NakajimaWest2017BJPS,McAlinnEtAl2017,ZhouNakajimaWest2014IJF}. However, much of this development relies 
on   batch processing via EM and MCMC methods; such methods  inherently prohibit both efficient sequential analysis and 
computational scaling with time series dimension. 
  
Our general setting builds on the multiscale innovations of
~\cite{BerryWest2018DCMM} that adapt  the {\em decouple-recouple} modeling strategy~\citep{GruberWest2016BA, ChenBanksWest2018,West2020Akaike} to dynamic latent factor DGLMs.   This enables both efficient sequential analysis, partially parallelized over series, as well as computational scaling that is {\em linear} in the number $N$ of linked time series. 
That said, the analysis strategy in~\cite{BerryWest2018DCMM} uses Monte Carlo simulation of the latent factor $\bphi_t$ from an assumed parametric form of external model $p(\bphi_t|\cM_0,\cD_{t-1})$, and this raises 
the computational load significantly; it implies that,  for forecasting and sequential inference, computations scale as $O(IN)$ where $I$ is the adopted Monte Carlo simulation sample size. In many current and emerging applications, such as very large-scale consumer sales forecasting~\citep[e.g.][]{BerryWest2018TSM} and large-scale dynamic network flow monitoring~\citep[e.g.][]{ChenETALdynets2016JASA,ChenBanksWest2018} among others, $N$ can be in the hundreds of thousands and more. Further,  sequential computations are often needed on time scales that obviate intense Monte Carlo.

\subsection{VBLB for Decoupled Univariate Latent Factor DGLMs} \label{section-vblb-univar-latentfactors}

A first advance arises in extending the VBLB analysis to each $\cM_i$ separately.  Suppose interest lies on just one series $y_{i,t}.$   The DGLM structure has lost the \lq\lq linear'' element due to the product term involving the latent factor process. That is,  $\lambda_{i,t} = \bh_{i,t}'\bgamma_{i,t}+ \bphi_t'\bbeta_{i,t}$  where both the factor $\bphi_t$ and its coefficient subvector $\bbeta_{i,t}$ are uncertain.  However, the VBLB analysis extends directly, as follows. 

In the independent external model $\cM_0$ at time $t-1,$ denote the mean vector and variance matrix of the latent factor $\bphi_t$ by 
$\bphi_t | \cD_{t-1} \sim [ \bb_t,\bB_t]$.  The latent factor is independent of $\btheta_{i,t} |\cD_t \sim [\ba_{i,t}, \bR_{i,t}]$ in $\cM_i$, so that moments can be directly computed using iterated expectations. Denote partitioned elements of the state moments by 
$$\ba_{i,t} = \begin{pmatrix} \ba_{\gamma,i,t} \\ \ba_{\beta,i,t}  \end{pmatrix} 
	\quad\textrm{and}\quad
	\bR_{i,t} =		\begin{pmatrix} \bR_{\gamma,i,t}& \bS_{i,t}\\  \bS_{i,t}' & \bR_{\beta,i,t} \end{pmatrix}. 
$$
Further, define $\tilde\bF_{i,t}$ as $\bF_{i,t}$ with the subvector $\bphi_t$ replaced by its mean $\bb_t,$ i.e., $\tilde\bF_{i,t} = (\bh_{i,t}',\bb_t')'.$ 
Then  $\lambda_{i,t} |\cD_t \sim [f_{i,t},q_{i,t}]$ with elements 
\begin{equation}\label{eqn-analytic-lf}
\begin{split}
f_{i,t} &= \tilde\bF_{i,t}'\ba_{i,t} = \bh_{i,t}' \ba_{\gamma,i,t}+ \bb_t'\ba_{\beta,i,t}, \\
q_{i,t} &=  \tilde\bF_{i,t}'\bR_{i,t}'\tilde\bF_{i,t} + \ba_{\beta,i,t}' \bB_t \ba_{\beta,i,t} + \tr(\bR_{\beta,i,t} \bB_t)
\end{split}
\end{equation}
where $\tr(\cdot)$ is the trace function. Further, the prior covariance vector 
$C(\btheta_{i,t},\lambda_{i,t}) = \bR_{i,t}\tilde\bF_{i,t}$ so  the implied adaptive vector in the VBLB update step is now $\bA_{i,t}=\bR_{i,t}\tilde\bF_{i,t}/q_{i,t}$.

The  VBLB-based filtering and forecasting analyses  apply with these modified expressions. Relative to the traditional DGLM analysis when $\bphi_t$ must be known, the equation for $f_{i,t}$,   the first term in $q_{i,t}$, and the expression for $\bA_{i,t}$ are as usual but with the latent $\bphi_t$ replaced by its mean $\bb_t.$ The final two terms in 
$q_{i,t}$ are new and adjust for  uncertainty about $\bphi_t$.
In multi-step  forecasting, the input moments of  state vector and latent factor will, of course, be replaced by relevant evolved values. 
 

\subsection{Cross-series Linkages of Latent Factor DGLMs}\label{recouple}
The series $i=1{:}N$ are related  through the recoupling role of the latent factor process.   Specifically, consider series $i$ and $j$ at time $t,$ with the summary information on series-specific state vectors given by 
 $\btheta_{i,t} \sim [\ba_{i,t}, \bR_{i,t}]$ in $\cM_i$ and  $\btheta_{j,t} \sim [\ba_{j,t}, \bR_{j,t}]$ in $\cM_j$, 
 conditionally independent given $\bphi_t.$    On marginalizing over $\bphi_t|\cD_t$ we have the implied means and variances for $\lambda_{i,t},\lambda_{j,t}$ of the previous Section. To add to this, the cross-series linkages induced are reflected in covariances 
\begin{equation} \label{eqn-joint-lfcovs}
q_{i,j,t} = C(\lambda_{i,t}, \lambda_{j,t}|\cD_t) = \ba_{\beta,i,t}' \bB_t \ba_{\beta,j,t}.
\end{equation}
Write $\blambda_t = (\lambda_{1,t},\ldots,\lambda_{N,t})'$.  Taken together over all series $i=1{:}N$, the implied 
mean vector and covariance matrix for this multivariate linear predictor vector is
\begin{equation} \label{eqn-joint-meanvar}
\blambda_t | \cD_t \sim [ \bff_t, \bQ_t ]
\end{equation}
where $\bff_t$ and $\bQ_t$ have elements defined in eqns.~(\ref{eqn-analytic-lf},\ref{eqn-joint-lfcovs}).  Again,  in multi-step ahead forecasting,  the input moments $\{ \ba_{i,t}, \bR_{i,t}, \bb_t, \bB_t \}$ 
 will  be replaced by   relevant evolved values.

\section{Copula Modelling for Efficient Multivariate Analysis} \label{analytic-inference}
 
 \subsection{Analytic Copula Construction \label{section:general-copula}} 
 
The traditional Bayesian analysis of DGLMS is semi-parametric, relying on first- and second-order moments of state vectors then conditioned via VB to define univariate marginal distributions for natural parameters in conjugate forms.   The interest here in multivariate structures-- in both dynamic latent factor models and related to multi-step forecasting--   focus questions about how to extend to specifications of joint distributions.   A copula approach is natural.   

For clarity, we abstract notation in this section. For some integer $p>0,$ consider a general vector $\by = (y_1,\ldots,y_p)'$ where each $y_j$ is drawn  from an exponential family distribution with  natural parameter $\mu_j$ and linear predictor $\lambda_j = g(\mu_j),$ and the $y_j$ are conditionally independent given the parameters. 
Write $\bmu=(\mu_1,\ldots,\mu_p)'$  and  $\blambda = (\lambda_1,\ldots,\lambda_p)'.$   The names of variables are unchanged relative to the prior section, but this general discussion drops the series $i,$ time $t$ indices for clarity and generality. This covers both multi-step forecasting and multiple series contexts.     Then, the vector of linear predictors has specified mean vector and variance matrix $\blambda \sim [ \bff,\bQ ]$ and the {\em marginal} distributions of the $\mu_j$ are constrained to be the unique conjugate forms  consistent with these moments. Denote the c.d.f.s of these conjugate  margins by $H_i(\mu_i),$ and their inverses (quantile functions) by $H_j^-(\cdot).$    

The copula approach defines a joint distribution $H(\bmu)$ via
\begin{equation}
\begin{split}
H(\bmu) 
&= G\left( G_1^{-}(H_1(\mu_1)), \dots, G_p^{-}(H_p(\mu_p))\right)
\end{split}
\end{equation}
based on a chosen multivariate copula distribution $G(\cdot)$ having univariate margins $G_i(\cdot)$ with quantile functions $G_i^{-}(\cdot).$   Key examples for $G(\cdot)$ in our DGLM contexts are distributions such that the transformed linear predictor vector $\blambda$ is multivariate normal or T. These allow for the full ranges of dependencies among the $\mu_j$ implied by those among the $\lambda_j,$ as well as being consistent with the view that prior and posterior distributions for the $\lambda_j$ will tend to be at least approximately symmetric.  
Importantly, the distribution $H(\bmu)$ has the exact margins $H_j(\mu_j),$ so that marginal inferences and implied marginal forecast distributions are precisely as in the univariate analyses.  The copula simply adds the parametric form incorporating dependencies given in $\bQ.$ 

The key use of the copula approach is in direct simulation of $\bmu$ to feed into simulations for forecasting.  Simply: 
\begin{itemize}\itemsep-3pt
	\item  generate a $p-$vector $\bz = (z_1,\ldots,z_p)'$ via $\bz \sim G(\bmu )$, noting that the sampled vector naturally reflects the dependencies defined in the underlying covariance matrix $\bQ$; 
	\item  for   $j=1{:}p,$ compute     $\mu_j= H_j^-(G_j(z_j))$, and then 
		sample $y_j|\mu_j$ from the exponential family distribution at that parameter value. 
\end{itemize}
Discussion  follows in examples below. 

\subsection{Illustration, Copula Choice and Computational Aspects} 

\begin{SCfigure}[][htbp!]
       \centering
	\includegraphics[height = 2.25in]{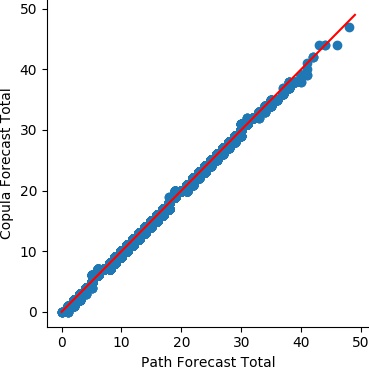}
	\caption{\label{fig-copula-path} Scatter plot of the empirical c.d.f.s of 100{,}000 samples from the 
	predictive distribution of the sum of 14 non-negative integer counts in a Poisson DGLM.  The copula method (vertical axis) 
	delivers basically the same results as the traditional DGLM analysis (horizontal axis) with under 1\% of the computational load. }
\end{SCfigure}

\paragraph{An Illustrative Vignette.} 
In advance of detailed case studies in the following sections,  an idea-fixing illustration comes from a daily sales forecasting application using a  Poisson DGLM with log link function. 
At a given time point, forecasts are made over $k=14$ steps ahead, so that $p=k=14$ and $\by$ is the vector of sales over each of the next 14 days. Figure \ref{fig-copula-path} 
relates to the implied forecast  distribution of $T= \bone'\by,$ the total sales over the next two weeks.  Computations are performed using (a)  the traditional DGLM analysis of Section~\ref{section-dglm-algos}, and (b)  the copula method in which $G(\bmu)$ is a multivariate log-normal distribution implied by a multivariate normal  $\blambda \sim N(\bff,\bQ).$  The Monte Carlo sample size is 
$100{,}000$   in each analysis.  The figure shows a scatter plot of the empirical c.d.f.s of the samples of $T.$  Given the discreteness of $T$, the resulting distributions are effectively indistinguishable.  It is also important to be reminded that the sampled marginal distributions of the $y_j$ are exactly equal.   Then, critically,  the computational time reduced from $292$ seconds in the traditional analysis to just $2.15$ seconds under the couple method. That is, basically the same results are achieved in about $0.74\%$ of the runtime.
\paragraph{Copula Model Choice.}   The choice of normal or T copula models is partly empirical and a matter for comparison and evaluation modulo the analysis goals. In multiple examples and experiments with real and synthetic data,  our experience is that the differences are typically very minimal in terms of the impact on forecast distributions.  That said, quantification of parameter uncertainty can be more impacted, since T models naturally define heavier tails in the assumed joint distribution $G(\bmu).$   In the context of a specific application, modelers may consider using T forms that can be expected to more adequately match-- at least theoretically-- the tails of the parameter prior and posterior distributions for the $\lambda_j$ under the relevant conjugate forms assumed in the VB steps of the DGLM analysis. Our examples in the following sections adopt normal copula models throughout; given the main focus of the paper on computational advances, moving to T copula forms adds negligibly to computational costs.

\paragraph{VB Optimization.}
One key computational detail is that the VB step to fix conjugate priors is really a main cost. This step evaluates the hyper-parameters of a conjugate prior with the moments of 
each $\lambda_j \sim [f_j, q_j]$. For example, in a Poisson DGLM with log link function,    the conjugate prior is $\mu_j \sim Ga(a_j, b_j)$ with 
 $ f_j =  E[\log(\mu_j)] = \gamma(a_j)-\log(b_j)$ and $ q_j = V[\log(\mu_t)] = \dot\gamma(a_j) $, and where $\gamma(\cdot)$ and $\dot\gamma(\cdot)$ are the digamma and trigamma functions, respectively.   These are typically, and most efficiently, solved for $(a_j,b_j)$ via Newton-Raphson. See this and other examples in chapter 14 of ~\cite{WestHarrison1997}. 
 However, in the traditional DGLM analysis in both multi-step forecasting per series and, more importantly, in latent factor models where these numerical solutions are needed for each of a set of Monte Carlo draws of latent factors,  this quickly becomes a dominant component of the computation at each time point. 
A novel strategy to reduce computations is to create a look-up table of values of the mapping $(f_j,q_j) \to (a_j,b_j)$ over a fine two-dimensional grid,  and then in analysis directly interpolate based on any realized $(f_j,q_j).$   In fact, this was already used in the example underlying Figure~\ref{fig-copula-path} in both the traditional and copula-based analysis. To assess its specific role,  we note that the  total time for drawing the $100{,}000$ samples from the path forecast distribution without look-up and interpolation is in fact around $1{,}880$ seconds, rather than the $292$ seconds with look-up and interpolation. Using the copula model with look-up and interpolation for the VB step reduced that to $2.15$ seconds, for an overall reduction to about $0.1\%$ of computational time relative to the traditional analysis.

%
\section{Multiscale Inference for Product Demand Forecasting} \label{section-salesforecasting}
\subsection{Application and Model Context}

One motivating context is in product demand and sales forecasting~\citep{BerryWest2018DCMM,BerryWest2018TSM}. Here we explore analysis of selected time series in this area using the  class of Dynamic Count Mixture Models (DCMMs) of~\cite{BerryWest2018DCMM}.  These involve linked Bernoulli and Poisson DGLMs as follows. In the notation of Section~\ref{section-dglm}, $y_{i,t}$ represents sales on day $t$ of a specific item $i$ in one supermarket.   Each model $\cM_i$ is a DCMM involving conditionally independent binary series $z_{i,t}$ as follows: 
\begin{equation} \label{eqn-dcmm}
\cM_i: \quad z_{i,t} \sim Ber(\pi_{i,t}) \quad \textrm{and}\quad y_{i,t} | z_{i,t}  = 
\begin{cases}
0, & \text{if } z_{i,t} = 0,\\
1 + x_{i,t}, \quad x_{i,t} \sim Po(\mu_{i,t}), & \textrm{if}\ z_{i,t} = 1.
\end{cases}
\end{equation}
The  $\pi_{i,t}$ and $\mu_{i,t}$ vary according to the dynamics of independent Bernoulli and Poisson DGLMs (each with natural link functions) respectively:
\begin{equation*}
\text{logit}(\pi_{i,t}) = \bF_{0, t}'\btheta_{0,t} \qquad \text{and} \qquad \log(\mu_{i,t}) = \bF_{+, t}' \btheta_{+, t}
\end{equation*}
with defined regression vectors and state vector evolutions. 
Our multi-scale model links across these item-level models to share information on a latent factor process $\bphi_t$ that is a subvector of both $\bF_{0,t}$ and $\bF_{+,t}.$ Inferences 
on $\bphi_t$ from an external model $\cM_0$ are passed down to each $\cM_i$ as discussed in Section \ref{section-dglm}. 

This example concerns daily sales of $N=2$ soda items in a single grocery store. 
The latent factor $\bphi_t$ includes day-of-week effects, holiday effects, and the aggregate sales forecast. This is inferred from the external model $\cM_0$, a traditional normal dynamic linear model (DLM) fitted to the log aggregate sales of all soda items. The DCMMs $\cM_i$ for individual soda items have item price and the latent factor $\bphi_t$ as predictors in both $\bF_{0,t}$ and $\bF_{+,t}.$    All model specifications are the same in each analysis. Finer modeling details are not relevant here as we are focused on the comparisons under the two computational approaches: the standard DGLM analysis that exploits recursive simulation into the future for step ahead forecasting, and the new partially \lq\lq analytic''  copula-based analysis.
We have data over several recent years. 
Models are trained on daily sales data from January 1st 2014 to June 30th 2015, and evaluated in multi-step predictions from July 1st 2015 to December 31st 2015. On each day in the evaluation period, analysis generates simulations of the $1{:}7$ day ahead forecast distributions for the future bivariate vector of sales outcomes.  

\begin{SCfigure}[][htbp!]
	\centering
	\includegraphics[height = 2.8in]{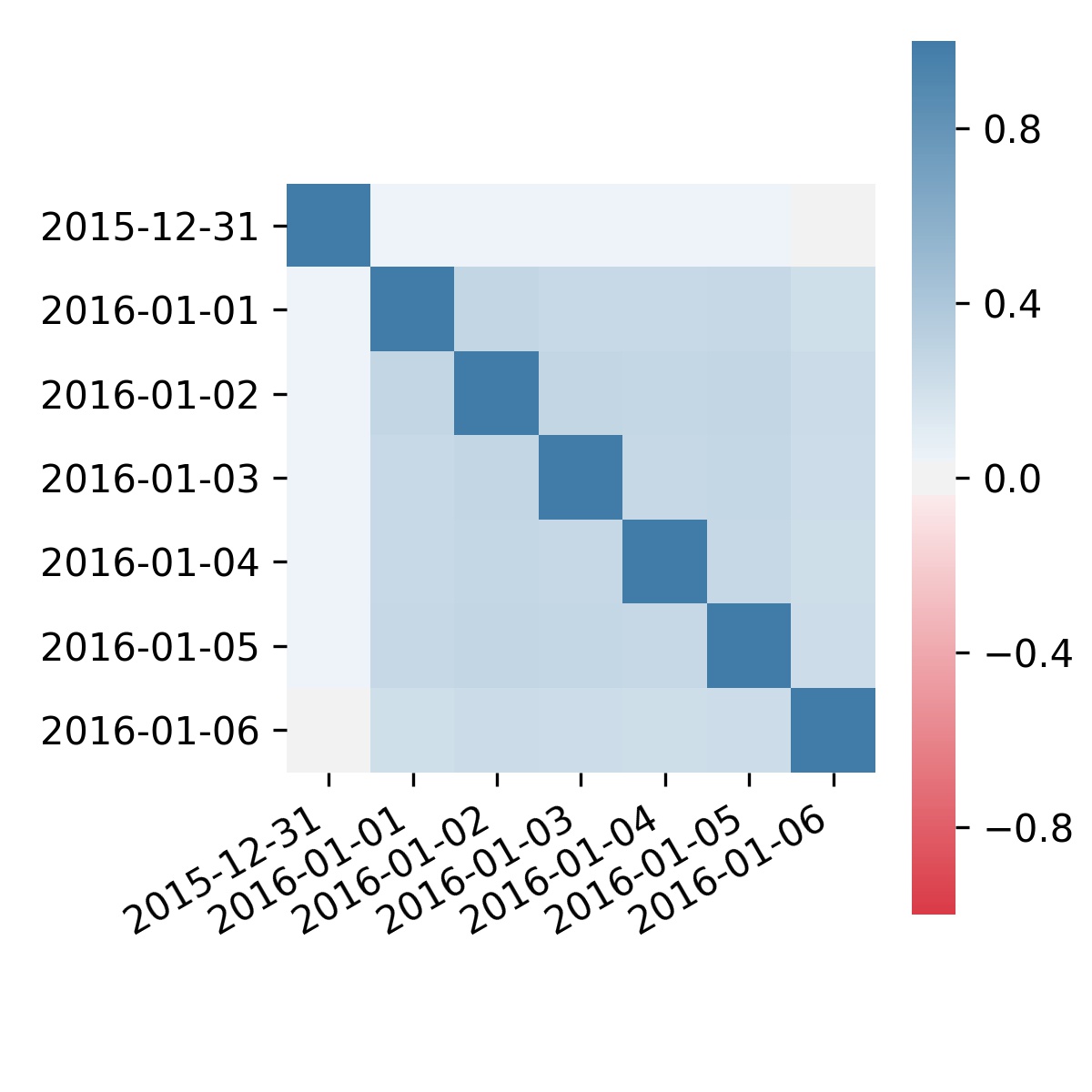}
	\caption{\label{fig-multiscale-corr}  Correlations in $\bQ_{1,T}$, the covariance matrix underlying the copula model for the $7-$day path forecast distribution $p(y_{1,T+1:T+t|\cD_T})$ of sales for item A $(i=1)$ made at time $T$ (December 30th 2015). 
	 Note (i) the typically positive correlations over the last 6 days; and (ii) as  the dynamic model includes a New Year's Eve holiday factor (as it does for other specific holidays),              December 31st is essentially uncorrelated with the other days.}
\end{SCfigure}

Insight into the nature of the dependence structure in multi-step forecasting is given by Figure~\ref{fig-multiscale-corr}.  This is drawn from the copula-based analysis of the 
full series up to time $T$ set at December 30th 2015. At that time, path forecasting defines the predictive distribution  $p(\by_{T+1:T+7}|\cD_T)$ for sales over the following 7 days. The figure illustrates correlations in $\bQ_{1, T}$, the covariance matrix underlying the copula model over the $7$ days for item A $(i=1).$ The path forecast distribution typically evidences strong correlations induced by those in the state vectors and latent factors; the patterns do, of course, vary in time.

%

The following sections give extracts of the analysis comparing the simulation-based  method of~\citep{BerryWest2018DCMM} and the copula method.  These include  filtered/online  estimates of  state vector elements and selected forecast accuracy and calibration metrics, all of which are expected to be substantially similar under the two approaches. More could be illustrated, but the central interest is on computational efficiency measures. Our experience and results in this study, as in others we have explored, demonstrate almost equivalent  multi-step ahead forecasting performance, while the analytic strategy using copul\ae\ provides most significant computational advantages.

\subsection{Example State Trajectories}   
\begin{figure}[htbp!]
	\centering
	\includegraphics[height = 2.8in]{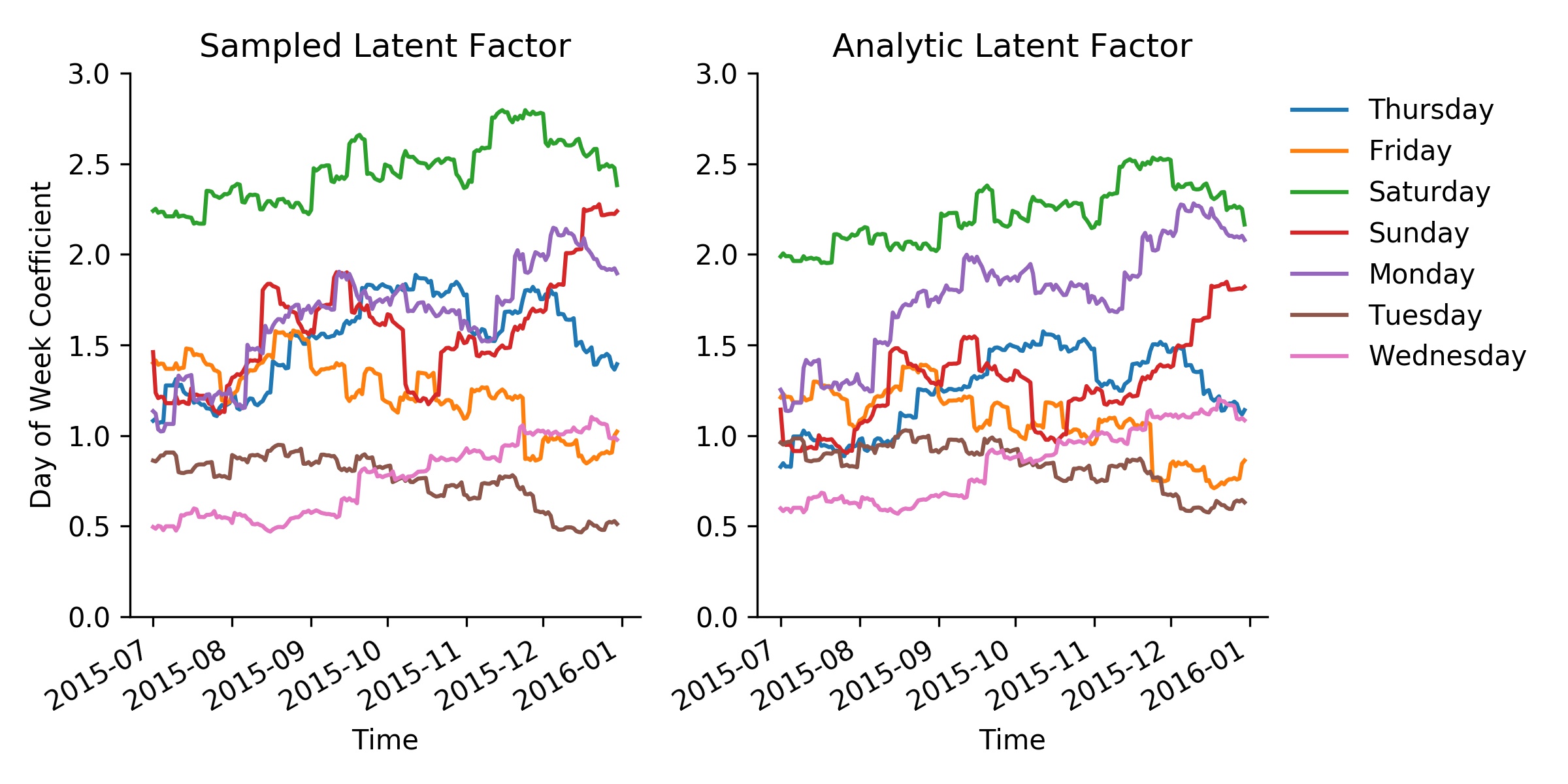}
	\caption{\label{fig-multiscale-dow-coef} Example from model for sales series A $(i=1)$.  The figures show online trajectories of filtered means of state vector coefficients for 
	the latent day-of-week factor, using the simulation-based analysis  (left) and the  copula-based \lq\lq analytic'' approach (right). }
\end{figure}

Figure~\ref{fig-multiscale-dow-coef} shows the trajectories of filtered means of the 7 day-of-week seasonality coefficients for item A (series $i=1),$ comparing those from each of the two computational methods.  These coefficients reflect how levels of item A respond to the inferred aggregate seasonal pattern in $\bphi_t$. The trajectories follow closely similar-- though not identical-- patterns using both computational methods. The consonance of trajectories is exhibited with other state vector elements, and in other examples we have explored; it is predicted as the marginal (series-specific) filtering updates are essentially the same in the two approaches, the results modified only based on the different approach to accounting for dependency across the item series. Importantly, the small differences in trajectories are rendered effectively irrelevant in forecasting once integrated with the corresponding posterior uncertainties (not shown in the figure).   

\subsection{Aspects of Forecast Assessment}
\begin{SCfigure}[][htbp!]
	\centering
	\includegraphics[width = 3.3in]{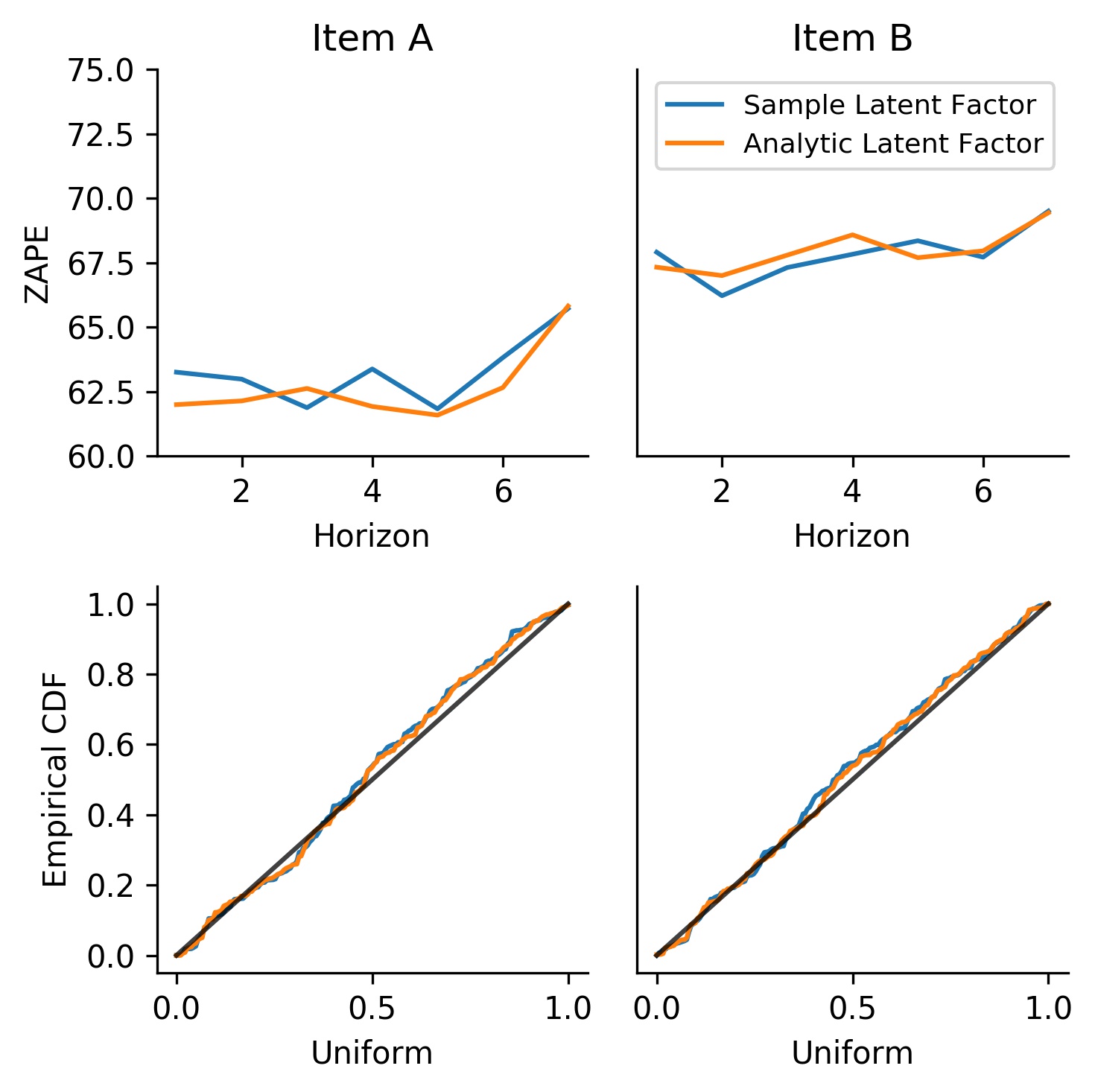}
	\caption{\label{fig-multiscale-zape-pit} Comparison of forecast metrics in forecasting sales of the two  item series over  the test period,   using the simulation-based analysis
	(\lq\lq Sampled Latent Factor'') and the copula method (\lq\lq  Analytic Latent Factor'').   The upper frames plot empirical mean ZAPE metrics as a function of forecast horizon over 1 to 7 days ahead, with values averaged over the full test period.  The lower frames display randomized PIT plots for the forecast weekly total sales for each item, again over the full test period.   }
\end{SCfigure}

\cite{BerryWest2018DCMM} develop a range of studies of forecast accuracy using multiple metrics for assessment. We choose two such metrics to illustrate the comparison between the two computational methods here. For point forecast accuracy, we consider the zero-adjusted percent error (ZAPE) that extends the industry standard percent error metric to allow for zero items in sales outcomes.  For a single day $t$ with point forecast $f$ and outcome $y$ on one item series, the ZAPE metric is
$ I(y= 0) f + I(y > 0)  |y - f| / y$. For series varying at relatively high levels, the probability of zero values  is small and the first term becomes negligible, so that the metric is close to the usual percent error; otherwise, the first term penalizes  larger point forecasts when zero sales occur.    In our data context,  the former is generally true for each of the two items, 
and we simply adopt medians of predictive distributions as point forecasts.    Then for each item, the average ZAPE values across the test period are evaluated.  Figure~\ref{fig-multiscale-zape-pit} exhibits results in forecasting across the test period with respect to multiple steps ahead over 1 to 7 days; evidently, the analytic copula-based method and the simulation-based method are effectively the same in terms of this metric.  This is consistent across other metrics including 
absolute deviation, square error, and others explored, bearing out the consistency of the two computational approaches in defining essentially equivalent forecast outcomes in these point forecast metrics. 

To go further in comparing forecast distributions,  we exploit the availability of full forecast distributions to explore distributional consonance between the two computational approaches.  The (randomized) probability integral transform (PIT)~\citep{BerryWest2018DCMM} is one of the most useful tools of forecast distributional assessment. PIT simply computes  the empirical distribution (over the test period) of the realized values of forecast c.d.f.s; then,  concordance with/deviations from  the quantiles of a uniform distribution can aid in assessing forecast model adequacy.    The randomized PIT values~\citep{Kolassa2016} adjust for the discreteness of predictive distributions so are used here. 
This is illustrated   with a focus on forecasting  total weekly sales,  i.e., for each item, the sum of sales over the next 7 days made repeatedly each day over the test period. 
Figure~\ref{fig-multiscale-zape-pit} provides further confirmation of the agreement between the analyses based on the two different computational methods, now comparing PIT plots. 
It also shows, as an aside, that the model seems to be well-calibrated with respect to this specific forecasting goal.

\subsection{Computation}
Table \ref{tab:multiscale-compute} shows the run times for updating and forecasting the two items over the $6$~month evaluation period. For this comparison, 
model updating at each time step used $I=200$ Monte Carlo samples of $\bphi_t$ in the standard analysis~\citep{BerryWest2018DCMM} that samples latent factors; the same sample size was using in both this and the new copula analysis in direct simulation of forecast distributions over $1{:}7$ days ahead at each time.   While a full practical analysis will use much larger sample sizes, this suffices for the comparison of compute time of main interest here.  The table gives running times broken  out for each of the filtering updates, the computation of marginal forecast distributions at each of the future 7 days, and that of the joint distribution, i.e., path forecast distribution, over the future 7 days at each time.  
 There is a significant computational improvement in using the analytic copula-based approach.  Compared to the sampled latent factor without interpolation for the VB step, there is a speedup of $810\times $ for model updating and $860\times$ for marginal forecasting.  The copula model analysis defines a $140\times$ speedup over standard path forecast simulation alone.
\begin{table}[h]
	\centering
	\begin{tabular}{lccc}
		& \multicolumn{1}{l}{{Update}} & \multicolumn{1}{l}{{Marginal Forecast}} & \multicolumn{1}{l}{{Path Forecast}} \\ 
		\multicolumn{1}{l}{{Sampled Latent Factor (No VB interp.):}}  & \multicolumn{1}{c}{1057.1}          & \multicolumn{1}{c}{1637.3}                     & \multicolumn{1}{c}{1823.4}                          \\ 
		\multicolumn{1}{l}{{Sampled Latent Factor:}}  & \multicolumn{1}{c}{202.6}          & \multicolumn{1}{c}{160.2}                     & \multicolumn{1}{c}{309.7}                          \\ 
		\multicolumn{1}{l}{{Analytic Latent Factor:}} & \multicolumn{1}{c}{1.3}            & \multicolumn{1}{c}{1.9}                       & \multicolumn{1}{c}{12.6}                           \\ 
	\end{tabular}
	\caption{Total compute time (seconds) for daily updating and multi-step forecasting performed over $6$ month period.}
	\label{tab:multiscale-compute}
\end{table}

It is also important to note that the copula model analysis is partially decoupled at each time point, so that computations   scale linearly in the number of item series $N.$  Hence the above example results can be extrapolated directly to larger numbers of series.   The new analysis decouples the computationally expensive VB operation from the number of 
Monte Carlo samples of $\bphi_t$ in the standard analysis, reducing the implied computations for that specific but key step from $O(IkN)$ to $O(kN)$ in forecasting over $k-$steps ahead each time in a model for $N$ series.  
After the VB step, the next most expensive operation is that for c.d.f. inversion in the copula analysis. This scales linearly with $I$ and $k$; in typical applied settings where $S\gg k$, the inverse c.d.f. run time dominates that for VB calculations. 


\section{Monitoring Network Flow Rates with Poisson DGLMs} \label{section-networkflow}
\subsection{Application and Model Context}
A second motivating context is that of time series of traffic flows on large-scale networks.  This is exemplified by e-commerce internet web traffic~\citep{ChenETALdynets2016JASA, ChenBanksWest2018} where network nodes within a commercial web site are industry-defined advertising categories and subcategories. 
Primary interest is in tracking and monitoring for unusual changes in flows,  with challenges of scaling in both time and dimension.  The above authors analyze 5-minute interval counts on networks with $N,$ the number of network node pairs, in the thousands or hundreds of thousands; streaming analysis at finer time scales and even larger networks is of interest.     Their approach uses  sets of decoupled Poisson DGLMs for individual node-node series $y_{i,t}$, 
and then recouples by mapping inferences to a dynamic gravity model for exploration of cross-series relationships. While effective, this is a start on a path towards more formal multivariate models incorporating latent processes that impact across the network and influence each series in stochastic ways; in addition to monitoring individual flows, we are interested in 
elucidating broader dynamics in network structure, overall traffic patterns and aggregate patterns on sub-networks,  and relative affinities between subsets of nodes, all of which call for dynamic latent factor modeling.  Building traditional hierarchical models  is not a path to pursue as it immediately limits scalability and fast sequential filtering. In contrast,  this is a perfect setting for multiscale modeling with the copula strategy. 
 
With $I$ network nodes,  we define models $\cM_{i,j}$ for the series of flows from each node $i$ to each node  $j$ in interval  $(t-1,t]$: 
\begin{equation}
\cM_{i,j}: \qquad y_{i,j,t} \sim Po(\mu_{i,j,t}), \qquad \lambda_{i,j,t} = \log(\mu_{i,j,t}) = \bF_{i,j,t}'\btheta_{i,j,t}
\end{equation}
where we extend the earlier development of latent factor models to  allow node-pair specific factors that nevertheless link across series. In detail: 
\begin{itemize} \itemsep-3pt
\item $\bF_{i,j,t}= (1,   \bphi_{i,j,t}')'$ involving a $2-$vector latent factor  $\bphi_{i,j,t} = ( \phi_{-,i,t}, \phi_{+,j,t})';$  
\item for all nodes $i=1{:}I,$  element $\phi_{-,i,t}$ is the log total flow out of category $i$ and $\phi_{+,i,t}$ is the log total flow into category $i$ over $(t-1,t].$ 
\item The evolution over time of  $\btheta_{i,j,t}$ is a vector random walk with levels of stochastic variation over time defined by discount factors, as usual. 
\end{itemize} 
The multiscale framework is completed with a set of external, or aggregate models $\cM_{0,i}$ for the $\phi_{-,i,t}$ and $\phi_{+,i,t}.$ In our case study below, these are traditional 
normal DLMs for the log counts out of/into each node in which the state vectors include local intercept and slope terms, i.e., defining 
local linear growth models (LLGMs) that are able to adapt to time-varying patterns over the day.  In other applications they could be defined by Poisson DGLMs or DCMMs, or other relevant model forms. 

The concept is that knowledge of the aggregate flow measures $ \bphi_{i,j,t}$ would provide information sharing across node-pairs and hence improved inference on local 
traffic patterns. The log outflows from node $i$ serve as a scaling factor, eliminating the need for empirically accounting for occupancy levels within nodes as was required in the analysis of~\cite{ChenETALdynets2016JASA, ChenBanksWest2018}.  Analytic inference with the latent factors is automatic, using a direct extension of the methods from Section~\ref{analytic-inference} and~\ref{recouple}. The copula construction acts to recouple the $\cM_{i,j}$ across nodes in joint forecasting $1-$step ahead, followed by the DGLM updates in each model.  Forecasting more steps ahead follows similarly, although this is not a main interest in network monitoring applications {\em per se}.

\subsection{Data, Monitoring and Dependence}
\begin{figure}[htbp!]
	\centering
	\includegraphics[height = 3.2in]{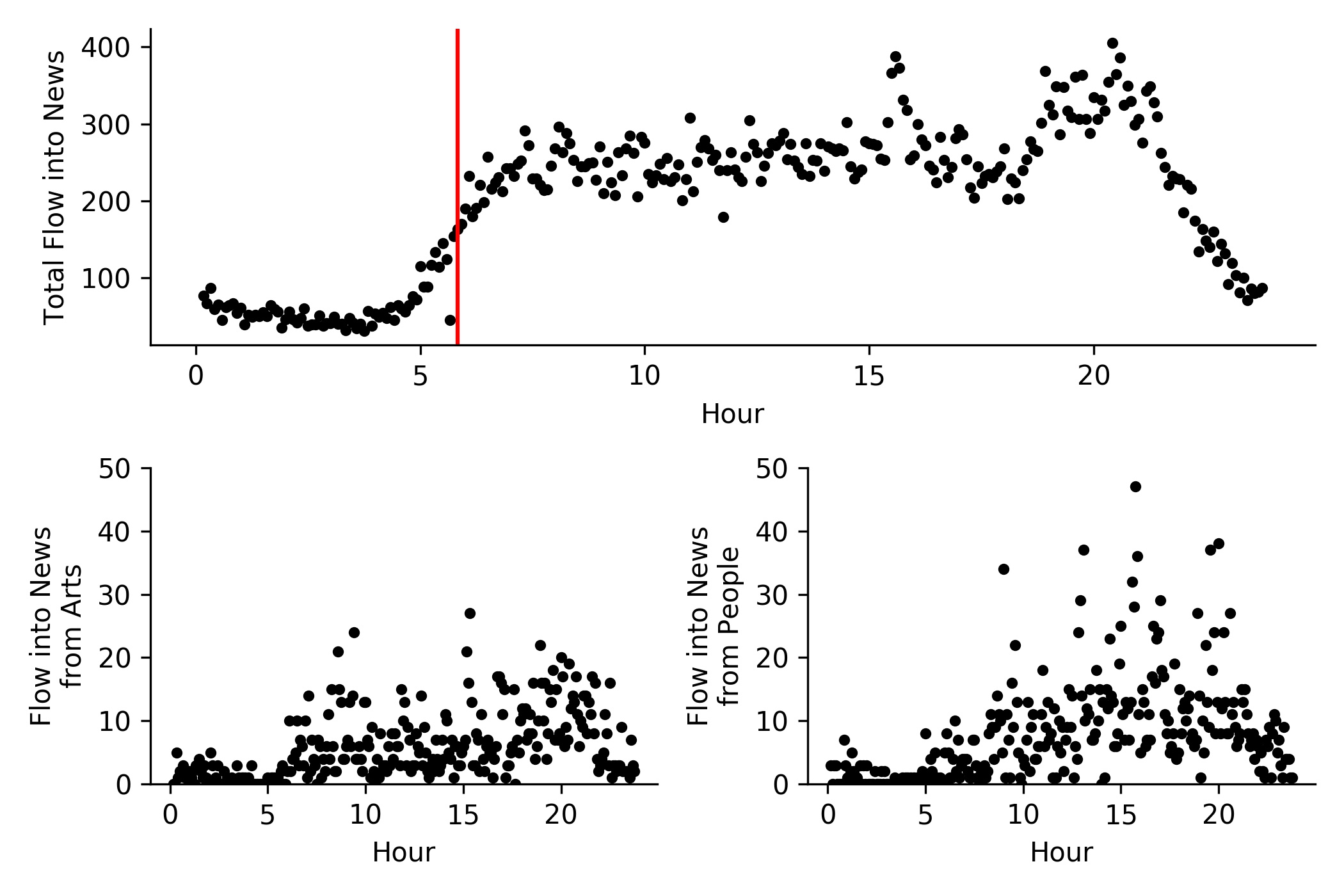}
	\caption{\label{fig-totalflow} The upper frame shows total flows into News over a $24$ hour period on September 17th 2015; the red line indicates the time of maximum correlation in the  copula model analysis. The lower frames show flows into News from Arts \& Entertainment (left) and from People \& Society (right). }
\end{figure}

We explore a subset of data from~\cite{ChenETALdynets2016JASA} concerning traffic on the Fox News website. Network nodes are advertising Adex Categories or subcategories such as Arts \& Entertainment, People \& Society, and News, etc. Every web page on the site is assigned to a category or subcategory.  Our example focuses on 
the most popular category, News, using $5-$minute interval data over a $24$ hour period on September 17th 2015. Figure~\ref{fig-totalflow} shows the total News inflows
as well as individual inflows from two specific nodes-- the Arts \& Entertainment node, and the People \& Society node. We select 11 nodes for  which average flows into News exceed $1$, defining 11 categories: Arts \& Entertainment, Games, Business \& Industrial, People \& Society, Law \& Government, Sports, Health, Autos \& Vehicles, Travel, Food \& Drink, and Science.
We build a copula-based dynamic latent factor model for the node-node time series on the resulting flow series. We compare aspects of the analysis with the results from the models of~\cite{ChenBanksWest2018}. Model and prior specifications are matched in detail, so that resulting inferences at the level of individual series  will differ based  on short-term predictive  model fit to the data evaluated on the  multivariate forecast distributions. 
We communicate aspects of model fit comparisons below, while continuing to be mainly concerned with comparing computational implications. 
\begin{SCfigure}[][htbp!]
	\centering
	\includegraphics[height = 3.2in]{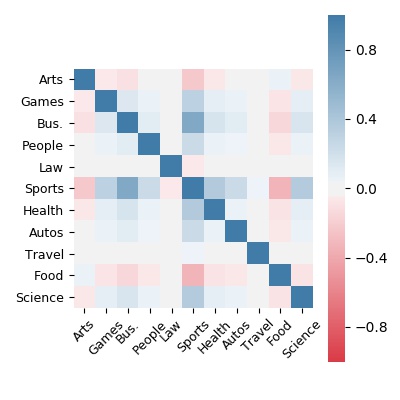}
	\caption{\label{fig-nodecorr} Correlations in $\bQ_{t}$, the covariance matrix underlying the copula model for $1-$step ahead forecasts of inflows to News, computed at the time immediately following  an anomalously low overall inflow to News.} 
\end{SCfigure}

Capturing the dependence among traffic flows patterns to accurately monitor and forecast their combined flows is one key reason to develop these latent factor models.  
To gain some insight, 
we mark (with the red line) in Figure~\ref{fig-totalflow} the time at which the copula model analysis indicates maximum correlation between any pair of nodes.  It turns out that this immediately follows a rather wild total inflow value, an observation that can be regarded as a potential outlier or a reason to explore and intervene in the model at that time.   In addition to simply exemplifying structure in the patterns of dependencies (which are, of course, time-varying), this speaks also to the interest in multiscale  modeling to begin, especially here for sequential monitoring.  The latent factor process for inflows experienced an anomalous event that has impact across all the node-specific inflows into News,  but is less apparent at that micro-level than in the external/aggregate level model that generates the latent factor processes.

\subsection{Aspects of Forecast Assessment}
Among other comparison metrics, Bayes' factors in favour/against  the copula-based multivariate factor model relative to the prior models of~\cite{ChenBanksWest2018} are natural considerations. 
Bayes' factors are special cases of the more general log predictive density ratio (LPDR) measures~\citep{NakajimaWest2013JBES, McAlinnEtAl2017} that compare forecast accuracy at 
$k-$steps ahead, evaluating 
$LPDR_{a/b}(k)	= \sum_{t=1:T} p_{a}(\by_{t+k} | \cD_t) / p_{b}(\by_{t+k} |\cD_t) $ over a period of time $t=1{:}T$ and where subscripts $a,b$ denote the two models being compared.  Here we have limited interest in more than $1-$step ahead forecasting; the traditional Bayes's factor  arises at $k=1,$  explicitly comparing (only) $1-$step ahead forecast accuracy under the two models.  This is a key focus when  monitoring for potential events impacting on some or all node flows is a main concern, as Bayes' factors underlie efficient sequential monitoring algorithms derived from Bayesian decision-theoretic perspectives~(\citealp{West1986a,West1986}, 
and~\citealp[chapter 11 of][]{WestHarrison1997}). 

\begin{figure}[htbp!]
	\centering
	\includegraphics[height = 2in]{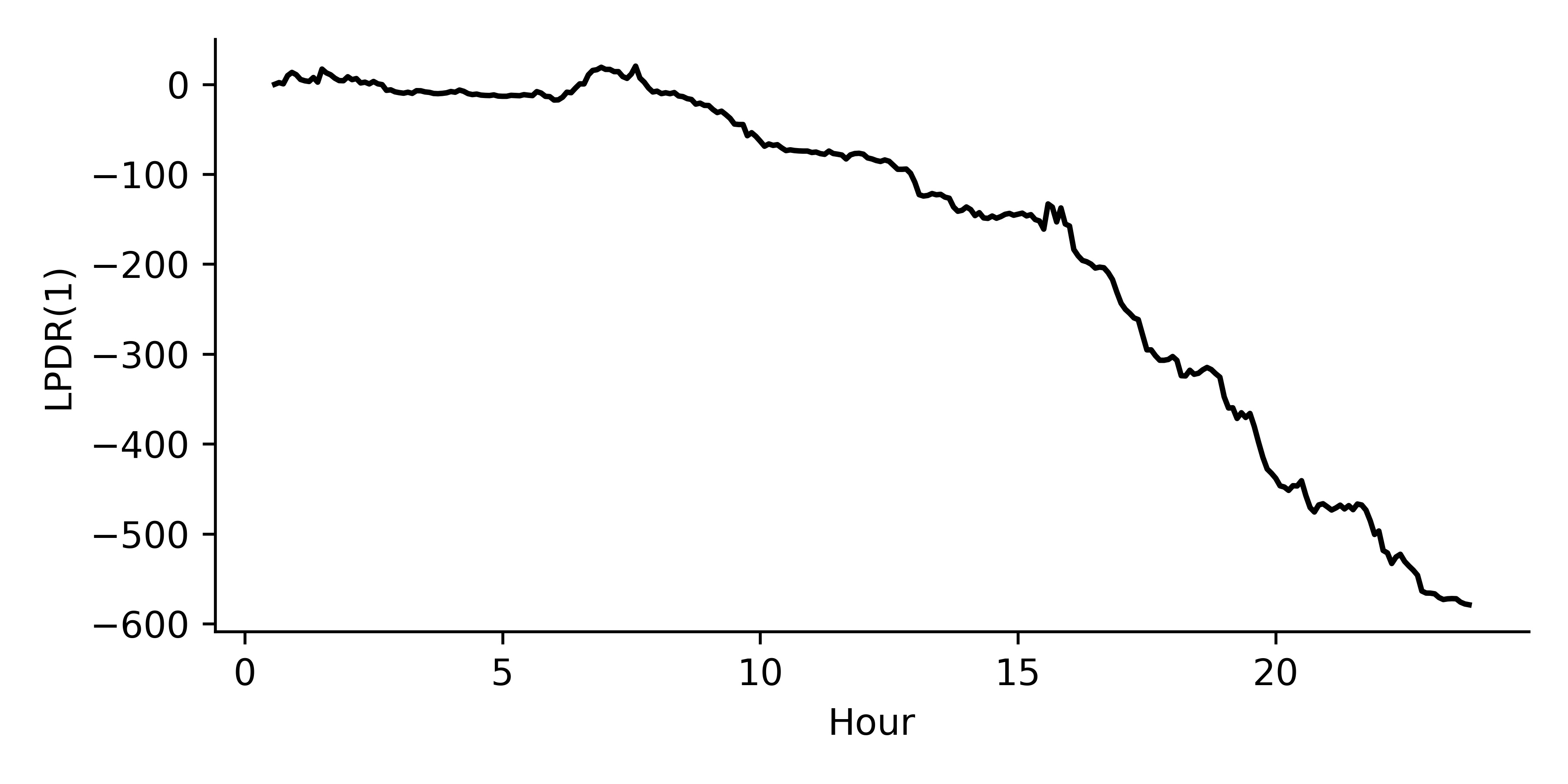}
	\caption{\label{fig-lpdr} Cumulative  $1-$step ahead log predictive density ratios (Bayes' factors) of decoupled Poisson DGLMs relative to the joint copula model. The flows are of visitors from $11$ other categories to the News category. The negative LPDR shows increasing evidence in favor of the copula model.}
\end{figure}

A key technical feature is that the sets of decoupled DGLMs of~\cite{ChenBanksWest2018} and the new multivariate copula-based models here all generate $1-$step forecast distributions in analytic forms based on the VBLB algorithms.  Hence, with model $a$ from~\cite{ChenBanksWest2018}
and model $b$ defined by the copula approach,  computation of $LPDR_{a/b}(1)$ is trivial. 
Figure~\ref{fig-lpdr} shows  cumulative $LPDR_{a/b}(1)$ evaluated from an initial  time point  after a short training period of $5$ observations, and running through   the remainder of the $24$ hour period. The negative LPDR reflects accumulated-- and increasingly strong--  evidence in favor of the joint copula model.  This arises from more relevant  characterization of dependencies among network flows and improved adaptation  to evolving data that results in substantially better  short-term forecast accuracy.   This is complimented by further 
evaluations of joint forecasts using randomized PIT measures (not shown) that confirm improved calibration under the copula-based model in forecasting cumulative node-node flows over short time periods; again, this arises due to appropriate characterization of dependencies across flows in the factor model, with analysis enabled via the copula construction. 

%
%
%
%

\subsection{Computation}
The fully decoupled DGLMs of~\cite{ChenBanksWest2018} are extremely computationally efficient for online filtering and forecasting. We have improved forecast performance by adding latent factors and jointly forecasting with a copula model. The additional state vector coefficients, and the need to generate joint samples of forecast distributions, add small computational overheads but they are counterbalanced by the speed-up from interpolation in the VB steps detailed earlier. 

Table~\ref{tab:network} shows run times for online updating and forecasting using $1{,}000$ samples from $1-$step ahead predictive distributions at each time point, repeated over $284$ time steps. Compared to the decoupled DGLMs without interpolation for the VB step, updating with the analytic, copula-based latent factor model was $3\times$  faster, while forecasting took just less than twice as long. All methods took under 10 seconds total to run the full analysis.

\begin{table}[h]
	\centering
	\begin{tabular}{lcc}
		& \multicolumn{1}{l}{{Update}} & \multicolumn{1}{l}{ {Forecast}} \\ 
		\multicolumn{1}{l}{{No Latent Factor (No VB interp.):}}       & \multicolumn{1}{c}{3.8}            & \multicolumn{1}{c}{4.1}              \\ 
		\multicolumn{1}{l}{{No Latent Factor:}}       & \multicolumn{1}{c}{0.5}            & \multicolumn{1}{c}{1.0}              \\ 
		\multicolumn{1}{l}{{Analytic Latent Factor:}} & \multicolumn{1}{c}{1.2}            & \multicolumn{1}{c}{7.8}              \\ 
	\end{tabular}
	\caption{Total compute time (seconds) for sequential updating and forecasting $1-$step ahead, for $11$ node-node networks flows over $284$ time steps.}
	\label{tab:network}
\end{table}

Copula-based updating is independent across flows, so run times scale linearly in the number of node pairs $N$. As in the previous application, a key operation in copula forecasting is the inverse c.d.f. transformation. The run time of these transformations scales linearly in $N$ and in the number of Monte Carlo samples desired. However, in larger networks, this quickly becomes much less of an issue as the Cholesky decomposition for multivariate normal or T copula sampling will become a more significant factor; network inference as the number of node pairs $N$ increases becomes increasingly computationally demanding in this aspect. Exploiting efficient numerical algorithms for Cholesky methods of large matrices, and potentially extending models to embed sparsity structure so enabling sparse matrix numerics, defines a set of interesting questions for potential next steps to address this.

\section{Summary Comments}
Motivated to scale-up computational methods for Bayesian sequential analysis in increasingly high-dimensional forecasting models, we have introduced the copula modeling approach for multi-step, multiscale and dynamic latent factor models that has proven effective in a number of contexts exemplified in this paper.  Examples demonstrate this in 
capturing dependence across time in joint/path  forecasting in DGLMs and in characterizing time-varying dependencies across series linked to latent factor processes. 
The theoretical basis extends existing DGLM analyses to a broad classes of forecasting and time series monitoring contexts, with coupled systems of binary, count and continuous time series outcomes in state-space settings. Partially analytic computations enable efficient evaluation of forecast parameters within and across series, with low computational overheads compared to standard univariate DGLMs.  This is enabled by  the decouple/recouple  concept, and defines  advances in both modeling and computation for Bayesian time series analysis and forecasting quite broadly.  

Examples in sales and product demand forecasting.  and in dynamic network traffic flow analysis,  demonstrate  the approach.  The sales forecasting  example achieves huge reduction in computational time relative to the existing approach that is much more heavily reliant on simulation.  In the network flow example, analysis based on existing decoupled models that are 
extended to integrate relevant multiscale factor components  improve forecasts and capture relevant dependencies with minimal computational overhead, and define more sensitive and adaptive models for flow monitoring and anomaly detection. Critically, both examples are in contexts where the new approach enables scaling in both time (data sampling rate) and dimension (number of related time series) with real practical opportunity.  

Some open questions arise in theoretical, computational and applied aspects.   There are potential theoretical research areas to explore in matching of copula model choices to known characteristics of implied distributions under the VB constrained conjugate-forms of univariate priors/posteriors in DGLMs.  Among technical questions with computational import, we have 
already noted above   the potential to exploit  efficient numerical algorithms for Cholesky methods of large matrices, and perhaps sparse matrix decomposition methods,  for multivariate normal and T copula distributions in models of increasing dimension.  Links to approaches in sparse factor 
modeling~\citep[e.g.][]{West2003,Carvalho2008,Lopes2008,Yoshida2010,NakajimaWest2013JFE,ZhouNakajimaWest2014IJF,Kaufman2019} in dynamic model contexts may be of interest here. Other technical themes include potential development of computational implementations that maximally exploit GPU, cloud and other distributed computational frameworks as part of the analysis. Decouple/recouple is at the heart of the computational advances, and is inherently geared towards partial parallelization. 
Another future research direction is to explore new settings for copula-based forecasting. One promising concept is to directly incorporate simultaneous outcomes of related series as predictors. In sales and product demand forecasting, for example, the sales from competing-- or compatible-- products can define useful predictors, potentially inducing dependence in a copula-based  model allied to that induced by latent factors in the current paper. This concept links to previous work with simultaneous predictors in normal dynamic models~\citep{ZhaoXieWest2016ASMBI, GruberWest2016BA, GruberWest2017ECOSTA} and can be expected to have broad applicability.

\appendix

\section{Appendix: Multi-step Forecasting in Copula DGLM Analysis} \label{appendix-dglm-path-forecast}
 
\subsection{Basic DGLM} 
In a DGLM for a series $y_t$ (dropping the series index $i$ here for clarity), interest lies in simulating $\by_{t+1:t+k} = (y_{t+1},\ldots,y_{t+k}|\cD_t)'$ from $p(\by_{t+1:t+k}|\cD_t)$ for some specified integer $k>0.$   The traditional recursive method in DGLMs has been summarized at the end of Section~\ref{section-dglm-algos}.  
The copula-based analysis is far more computationally efficient as discussed and exemplified in the paper.    The path of linear predictors is 
 $\blambda_{t+1:t+k} =  (\lambda_{t+1},\ldots,\lambda_{t+k})'$ where $\lambda_h = g(\mu_{t+h}) = \bF_h'\btheta_h$ for all $h,$  with corresponding natural parameters  $\bmu_{t+1:t+k} =  (\mu_{t+1},\ldots,\mu_{t+k})'.$   With Monte Carlo sample size $I,$ the path forecasting algorithm in the copula-based analysis is as follows. 
  
\begin{enumerate} \itemsep-3pt
\item Evaluate the moments of  $ \blambda_{t+1:t+k}|\cD_t \sim [ \bff_{t+1:t+k}, \bQ_{t+1:t+k}]. $ 
\item Identify the $k-$variate copula distribution $G(\cdot)$ for $\bmu_{t+1:t+k}$ consistent with these moments and having univariate margins $G_{t+h}(\cdot)$ for $h=1{:}k.$ 
\item Evaluate the hyper-parameters of the conjugate marginal prior distributions  $H_{t+h}(\cdot)$ for each $\mu_{t+h}|\cD_t.$
\item For $s=1{:}I,$  sample as follows: 
\begin{itemize}  
	\item  Simulate $ \bz_{t+1:t+k}^{(s)}  =  (z_{t+1}^{(s)},\ldots,z_{t+k}^{(s)})'$ from $ G(\cdot).$  
	\item  For $h = 1{:}k,$
	\begin{itemize}  
		\item calculate $ \mu_{t+h} = H_{t+h}^{-} \left( G_{t+h} (z_{t+h}^{(s)}) \right),$ then 
		\item simulate $ y_{t+h}^{(s)} $ from the exponential family distribution $p(y_{t+h} | \mu_{t+h}^{(s)}).$
	\end{itemize} 
	\item Save $ \by_{t+1:t+k}^{(s)} = (y_{t+1}^{(s)},\ldots,y_{t+k}^{(s)})'.$
 \end{itemize} 
\end{enumerate}
Initialization in Step 1 above involves computing moments of the $k-$vector of linear predictors; this follows trivially from standard linear model theory.  
Beginning with the current posterior moments for the state vector  $\btheta_t | \cD_t = [ \bm_t,\bC_t]$, 
the moments of $\btheta_{t+1:t+k}|\cD_t$ are trivially computed by recursing through the state evolution 
 equation over $t+1,\ldots,t+k$. This is standard in all linear models as detailed in~\citealp{WestHarrison1997} (Theorem 4.2 of section 4.4). 
 For each $h,j=1{:}k,$ the quantities are mean vectors
$ \ba_t(h) = E( \btheta_{t+h} | \cD_t),$ marginal variance matrices $\bR_t(h) = V( \btheta_{t+h} | \cD_t)$ and, for $j\ne h,$ 
covariance matrices $ \bC_t(h,j) = C( \btheta_{t+h}, \btheta_{t+j}| \cD_t)$.
For the implied linear predictor moments,  the mean vector $\bff_{t+1:t+k}$ has elements  
$f_t(h)  = \bF_{t+h}' \ba_t(h)$,   while the variance matrix $\bQ_{t+1:t+k}$ has diagonal elements  
$q_t(h)   = \bF_{t+h}'\bR_t(h) \bF_{t+h} $ 
and off-diagonal elements
$q_t(h,j)   = \bF_{t+h}'\bC_t(h,j) \bF_{t+j}$. 

\subsection{Latent Factor DGLM}

The above requires only modest changes in detail  to extend to the dynamic latent factor DGLM setting of 
Section~\ref{section-vblb-univar-latentfactors}. The expressions above are modified in that the $\bF_\ast$ vectors have latent factor components from $\bphi_\ast$  
substituted by their current forecast means from the external model $\cM_0,$ with some correction of variances and covariances
 to account for the uncertainty about the latent factor process into the future. 

Under $\cM_0$ denote step-ahead forecast moments of the latent factor process by 
$$
\bphi_{t+h} | \cD_t \sim \left[ \bb_t(h), \bB_t(h) \right] \quad\textrm{and}\quad 
C\left(\bphi_{t+h}, \bphi_{t+j} | \cD_t \right) = \bB_t(h,j) \ \textrm{for}\ j\ne h.
$$
Extend the notation of Section~\ref{section-vblb-univar-latentfactors} as follows. Define $\tilde\bF_{t+h}$ to be $\bF_{t+h}$ with the subvector $\bphi_{t+h}$ replaced by its forecast mean $\bb_t(h)$. Denote by $\ba_{\beta,t}(h)$ the subvector of $\ba_t(h)$ related to the latent factor state vector coefficients, by $\bR_{\beta,t}(h)$ the corresponding variance matrix block of $\bR_t(h),$ and by
$\bC_{\beta,t}(h,j)$ the corresponding covariance matrix block of each $\bC_t(h,j).$
Then  $\bff_{t+1:t+k}$ has elements  
$f_t(h)  = \tilde\bF_{t+h}' \ba_t(h)$ while $\bQ_{t+1:t+k}$ has diagonal elements
$$q_t(h)   = \tilde\bF_{t+h}'\bR_t(h) \tilde\bF_{t+h} + \ba_{\beta,t}(h)' \bB_t(h) \ba_{\beta,t}(h) + \tr(\bR_{\beta,t}(h) \bB_t(h)). $$
and  off-diagonal elements 
 $$q_t(h,j)   = \tilde\bF_{t+h}'\bC_t(h,j) \tilde\bF_{t+j} + \ba_{\beta,t}(h)' \bB_t(h,j) \ba_{\beta,t}(j) + \tr(\bC_{\beta,t}(h,j) \bB_t(h,j)). $$

\bigskip
\begin{center}
	{\large \textbf{SUPPLEMENTARY MATERIAL}}
\end{center}

\noindent {\blu\bf PyBATS}: A Python package for {\underbar B}ayesian {\underbar A}nalysis of {\underbar T}ime {\underbar S}eries and Bayesian forecasting using the general class of DGLM state-space models. Multivariate DGLM analysis is incorporated with dynamic latent factors using the multiscale approach. Includes models for count time series as developed in~\cite{BerryWest2018DCMM}, ~\cite{BerryWest2018TSM}, and the current paper.

-- PyBATS code repository: \href{https://github.com/lavinei/pybats}{https://github.com/lavinei/pybats}

-- PyBATS documentation:  \href{https://lavinei.github.io/pybats/}{https://lavinei.github.io/pybats/}

-- PyBATS examples: \href{https://github.com/lavinei/pybats/tree/master/examples}{https://github.com/lavinei/pybats/tree/master/examples}



\bibliography{LavineCronWest2019}

\begin{thebibliography}{}

\bibitem[\protect\citeauthoryear{Aktekin, Polson, and Soyer}{Aktekin
  et~al.}{2018}]{Aktekin2018}
Aktekin, T., N.~Polson, and R.~Soyer (2018).
\newblock Sequential {B}ayesian analysis of multivariate count data.
\newblock {\em Bayesian Analysis\/}~{\em 13}, 385--409.

\bibitem[\protect\citeauthoryear{Berry, Helman, and West}{Berry
  et~al.}{2019}]{BerryWest2018TSM}
Berry, L.~R., P.~Helman, and M.~West (2019).
\newblock Probabilistic forecasting of heterogeneous consumer transaction-sales
  time series.
\newblock {\em International Journal of Forecasting\/}~{\em 36}, 552--569.

\bibitem[\protect\citeauthoryear{Berry and West}{Berry and
  West}{2020}]{BerryWest2018DCMM}
Berry, L.~R. and M.~West (2020).
\newblock Bayesian forecasting of many count-valued time series.
\newblock {\em Journal of Business and Economic Statistics {\rm (in press)}\/}.

\bibitem[\protect\citeauthoryear{Cargnoni, M{\H u}ller, and West}{Cargnoni
  et~al.}{1997}]{Cargnoni1997}
Cargnoni, C., P.~M{\H u}ller, and M.~West (1997).
\newblock Bayesian forecasting of multinomial time series through conditionally
  {G}aussian dynamic models.
\newblock {\em Journal of the American Statistical Association\/}~{\em 92},
  640--647.

\bibitem[\protect\citeauthoryear{Carvalho, Johannes, Lopes, and
  Polson}{Carvalho et~al.}{2010}]{Carvalho2010}
Carvalho, C.~M., M.~S. Johannes, H.~F. Lopes, and N.~G. Polson (2010).
\newblock Particle learning and smoothing.
\newblock {\em Statistical Science\/}~{\em 25}, 88--106.

\bibitem[\protect\citeauthoryear{Carvalho, Lopes, and Aguilar}{Carvalho
  et~al.}{2011}]{Carvalho11}
Carvalho, C.~M., H.~F. Lopes, and O.~Aguilar (2011).
\newblock Dynamic stock selection strategies: {A} structured factor model
  framework (with discussion).
\newblock In J.~M. Bernardo, M.~J. Bayarri, J.~O. Berger, A.~P. Dawid,
  D.~Heckerman, A.~F.~M. Smith, and M.~West (Eds.), {\em Bayesian Statistics
  9}, pp.\  69--90. Oxford University Press, Oxford UK.

\bibitem[\protect\citeauthoryear{Carvalho, Lucas, Wang, Chang, Nevins, and
  West}{Carvalho et~al.}{2008}]{Carvalho2008}
Carvalho, C.~M., J.~E. Lucas, Q.~Wang, J.~Chang, J.~R. Nevins, and M.~West
  (2008).
\newblock High-dimensional sparse factor modelling-- {A}pplications in gene
  expression genomics.
\newblock {\em Journal of the American Statistical Association\/}~{\em 103},
  1438--1456.

\bibitem[\protect\citeauthoryear{Chen, Banks, and West}{Chen
  et~al.}{2019}]{ChenBanksWest2018}
Chen, X., D.~Banks, and M.~West (2019).
\newblock Bayesian dynamic modeling and monitoring of network flows.
\newblock {\em Network Science\/}~{\em 7}, 292--318.

\bibitem[\protect\citeauthoryear{Chen, Irie, Banks, Haslinger, Thomas, and
  West}{Chen et~al.}{2018}]{ChenETALdynets2016JASA}
Chen, X., K.~Irie, D.~Banks, R.~Haslinger, J.~Thomas, and M.~West (2018).
\newblock Scalable {B}ayesian modeling, monitoring and analysis of dynamic
  network flow data.
\newblock {\em Journal of the American Statistical Association\/}~{\em 113},
  519--533.

\bibitem[\protect\citeauthoryear{Del~Negro and Otrok}{Del~Negro and
  Otrok}{2008}]{DelNegro2008}
Del~Negro, M. and C.~M. Otrok (2008).
\newblock Dynamic factor models with time-varying parameters: {M}easuring
  changes in international business cycles.
\newblock Staff Report 326, New York Federal Reserve.

\bibitem[\protect\citeauthoryear{Ferreira, Gamerman, and Migon}{Ferreira
  et~al.}{1997}]{FerreiraGamermanMigon1997}
Ferreira, M. A.~R., D.~Gamerman, and H.~S. Migon (1997).
\newblock Bayesian dynamic hierarchical models: {C}ovariance matrices
  estimation and nonnormality.
\newblock {\em Brazilian Journal of Probability and Statistics\/}~{\em 11},
  67--79.

\bibitem[\protect\citeauthoryear{Ferreira, West, Lee, and Higdon}{Ferreira
  et~al.}{2006}]{Ferreira2006}
Ferreira, M. A.~R., M.~West, H.~K.~H. Lee, and D.~M. Higdon (2006).
\newblock Multiscale and hidden resolution time series models.
\newblock {\em Bayesian Analysis\/}~{\em 2}, 294--314.

\bibitem[\protect\citeauthoryear{Gamerman and Migon}{Gamerman and
  Migon}{1993}]{GamermanMigon1993}
Gamerman, D. and H.~S. Migon (1993).
\newblock Dynamic hierarchical models.
\newblock {\em Journal of the Royal Statistical Society (Series B:
  Methodological)\/}~{\em 55}, 629--642.

\bibitem[\protect\citeauthoryear{Gruber and West}{Gruber and
  West}{2016}]{GruberWest2016BA}
Gruber, L.~F. and M.~West (2016).
\newblock {GPU}-accelerated {B}ayesian learning in simultaneous graphical
  dynamic linear models.
\newblock {\em Bayesian Analysis\/}~{\em 11}, 125--149.

\bibitem[\protect\citeauthoryear{Gruber and West}{Gruber and
  West}{2017}]{GruberWest2017ECOSTA}
Gruber, L.~F. and M.~West (2017).
\newblock Bayesian forecasting and scalable multivariate volatility analysis
  using simultaneous graphical dynamic linear models.
\newblock {\em Econometrics and Statistics\/}~{\em 3}, 3--22.

\bibitem[\protect\citeauthoryear{Kastner, Fr{\H u}hwirth-Schnatter, and
  Lopes}{Kastner et~al.}{2017}]{kastner2017}
Kastner, G., S.~Fr{\H u}hwirth-Schnatter, and H.~F. Lopes (2017).
\newblock Efficient {Bayesian} inference for multivariate factor stochastic
  volatility models.
\newblock {\em Journal of Computational and Graphical Statistics\/}~{\em 26},
  905--917.

\bibitem[\protect\citeauthoryear{Kaufmann and Schumacher}{Kaufmann and
  Schumacher}{2019}]{Kaufman2019}
Kaufmann, S. and C.~Schumacher (2019).
\newblock Bayesian estimation of sparse dynamic factor models with
  order-independent and ex-post mode identification.
\newblock {\em Journal of Econometrics\/}~{\em 210}, 116--134.

\bibitem[\protect\citeauthoryear{Kolassa}{Kolassa}{2016}]{Kolassa2016}
Kolassa, S. (2016).
\newblock Evaluating predictive count data distributions in retail sales
  forecasting.
\newblock {\em International Journal of Forecasting\/}~{\em 32}, 788--803.

\bibitem[\protect\citeauthoryear{Lopes and Carvalho}{Lopes and
  Carvalho}{2007}]{LopesCarvalho07}
Lopes, H.~F. and C.~M. Carvalho (2007).
\newblock Factor stochastic volatility with time varying loadings and {M}arkov
  switching regimes.
\newblock {\em Journal of Statistical Planning and Inference\/}~{\em 137},
  3082--3091.

\bibitem[\protect\citeauthoryear{Lopes, Salazar, and Gamerman}{Lopes
  et~al.}{2008}]{Lopes2008}
Lopes, H.~F., E.~Salazar, and D.~Gamerman (2008).
\newblock Spatial dynamic factor analysis.
\newblock {\em Bayesian Analysis\/}~{\em 3}, 759--792.

\bibitem[\protect\citeauthoryear{McAlinn, Aastveit, Nakajima, and West}{McAlinn
  et~al.}{2020}]{McAlinnEtAl2017}
McAlinn, K., K.~A. Aastveit, J.~Nakajima, and M.~West (2020).
\newblock Multivariate {B}ayesian predictive synthesis in macroeconomic
  forecasting.
\newblock {\em Journal of the American Statistical Association {\rm (in
  press)}\/}.

\bibitem[\protect\citeauthoryear{McAlinn and West}{McAlinn and
  West}{2019}]{McAlinnWest2017bpsJOE}
McAlinn, K. and M.~West (2019).
\newblock Dynamic {B}ayesian predictive synthesis in time series forecasting.
\newblock {\em Journal of Econometrics\/}~{\em 210}, 155--169.

\bibitem[\protect\citeauthoryear{Migon and Harrison}{Migon and
  Harrison}{1985}]{migon1985application}
Migon, H.~S. and P.~J. Harrison (1985).
\newblock An application of non-linear {B}ayesian forecasting to television
  advertising.
\newblock In J.~M. Bernardo, M.~H. DeGroot, D.~V. Lindley, and A.~F.~M. Smith
  (Eds.), {\em Bayesian {S}tatistics 2}, pp.\  681--696. North-Holland,
  Amsterdam, and Valencia University Press, Valencia.

\bibitem[\protect\citeauthoryear{Nakajima and West}{Nakajima and
  West}{2013a}]{NakajimaWest2013JBES}
Nakajima, J. and M.~West (2013a).
\newblock Bayesian analysis of latent threshold dynamic models.
\newblock {\em Journal of Business and Economic Statistics\/}~{\em 31},
  151--164.

\bibitem[\protect\citeauthoryear{Nakajima and West}{Nakajima and
  West}{2013b}]{NakajimaWest2013JFE}
Nakajima, J. and M.~West (2013b).
\newblock Bayesian dynamic factor models: {L}atent threshold approach.
\newblock {\em Journal of Financial Econometrics\/}~{\em 11}, 116--153.

\bibitem[\protect\citeauthoryear{Nakajima and West}{Nakajima and
  West}{2017}]{NakajimaWest2017BJPS}
Nakajima, J. and M.~West (2017).
\newblock Dynamics and sparsity in latent threshold factor models: {A} study in
  multivariate {EEG} signal processing.
\newblock {\em Brazilian Journal of Probability and Statistics\/}~{\em 31},
  701--731.

\bibitem[\protect\citeauthoryear{Triantafyllopoulos}{Triantafyllopoulos}{2009}]{Triantafyllopoulos2009}
Triantafyllopoulos, K. (2009).
\newblock Inference of dynamic generalized linear models: {O}n-line computation
  and appraisal.
\newblock {\em International Statistical Review\/}~{\em 77}, 430--450.

\bibitem[\protect\citeauthoryear{West}{West}{1986}]{West1986a}
West, M. (1986).
\newblock Bayesian model monitoring.
\newblock {\em Journal of the Royal Statistical Society (Series B)\/}~{\em 48},
  70--78.

\bibitem[\protect\citeauthoryear{West}{West}{2003}]{West2003}
West, M. (2003).
\newblock Bayesian factor regression models in the \lq\lq large p, small n"
  paradigm.
\newblock In J.~M. Bernardo, M.~J. Bayarri, J.~O. Berger, A.~P. David,
  D.~Heckerman, A.~F.~M. Smith, and M.~West (Eds.), {\em Bayesian Statistics
  7}, pp.\  723--732. Oxford University Press.

\bibitem[\protect\citeauthoryear{West}{West}{2020}]{West2020Akaike}
West, M. (2020).
\newblock Bayesian forecasting of multivariate time series: {S}calability,
  structure uncertainty and decisions (with discussion).
\newblock {\em Annals of the Institute of Statistical Mathematics\/}~{\em 72},
  1--44.

\bibitem[\protect\citeauthoryear{West and Harrison}{West and
  Harrison}{1986}]{West1986}
West, M. and P.~J. Harrison (1986).
\newblock Monitoring and adaptation in {B}ayesian forecasting models.
\newblock {\em Journal of the American Statistical Association\/}~{\em 81},
  741--750.

\bibitem[\protect\citeauthoryear{West and Harrison}{West and
  Harrison}{1997}]{WestHarrison1997}
West, M. and P.~J. Harrison (1997).
\newblock {\em Bayesian Forecasting and Dynamic Models\/} (2nd ed.).
\newblock Springer.

\bibitem[\protect\citeauthoryear{West, Harrison, and Migon}{West
  et~al.}{1985}]{West1985a}
West, M., P.~J. Harrison, and H.~S. Migon (1985).
\newblock Dynamic generalised linear models and {B}ayesian forecasting (with
  discussion).
\newblock {\em Journal of the American Statistical Association\/}~{\em 80},
  73--97.

\bibitem[\protect\citeauthoryear{Yoshida and West}{Yoshida and
  West}{2010}]{Yoshida2010}
Yoshida, R. and M.~West (2010).
\newblock Bayesian learning in sparse graphical factor models via annealed
  entropy.
\newblock {\em Journal of Machine Learning Research\/}~{\em 11}, 1771--1798.

\bibitem[\protect\citeauthoryear{Zhao, Xie, and West}{Zhao
  et~al.}{2016}]{ZhaoXieWest2016ASMBI}
Zhao, Z.~Y., M.~Xie, and M.~West (2016).
\newblock Dynamic dependence networks: {F}inancial time series forecasting and
  portfolio decisions (with discussion).
\newblock {\em Applied Stochastic Models in Business and Industry\/}~{\em 32},
  311--339.

\bibitem[\protect\citeauthoryear{Zhou, Nakajima, and West}{Zhou
  et~al.}{2014}]{ZhouNakajimaWest2014IJF}
Zhou, X., J.~Nakajima, and M.~West (2014).
\newblock Bayesian forecasting and portfolio decisions using dynamic dependent
  sparse factor models.
\newblock {\em International Journal of Forecasting\/}~{\em 30}, 963--980.

\end{thebibliography}
\bibliographystyle{Chicago} 

\end{document}